\documentclass[
	a4paper, 
	10pt, 
]{LTJournalArticle}

\usepackage[utf8]{inputenc} 
\usepackage[T1]{fontenc}    
\usepackage[english]{babel}
\usepackage{hyperref}       
\usepackage{url}            
\usepackage{booktabs}       
\usepackage{amsfonts}       
\usepackage{nicefrac}       
\usepackage{microtype}      
\usepackage{lipsum}         
\usepackage{graphicx}
\usepackage{doi}
\usepackage{multirow}
\usepackage{makecell}
\usepackage{gensymb}
\usepackage{dsfont}
\usepackage{amsmath}
\usepackage{bm}
\usepackage{float}
\usepackage{multicol}
\usepackage{setspace}

\usepackage[acronym]{glossaries}
\usepackage{cleveref}       
\usepackage{lineno}

\usepackage[
backend=biber,
style=authoryear,
natbib=true,
date=year,
maxnames=2,
maxbibnames=99,
isbn=false]{biblatex}

\newacronym{bs}{BS}{Brier Score}
\newacronym{crps}{CRPS}{Continuous Ranked Probability Score}
\newacronym{drn}{DRN}{Distributional Regression Network}
\newacronym{ecc}{ECC}{Ensemble Copula Coupling}
\newacronym{es}{ES}{Energy Score}
\newacronym{esgm}{ESGM}{Energy Score Generative Model}
\newacronym{fmap}{FMAP}{Flow MAtching Postprocessing}
\newacronym{gan}{GAN}{Generative Adversarial Network}
\newacronym{ode}{ODE}{Ordinary Differential Equation}
\newacronym{nudft}{NUDFT}{Non-Uniform Discrete Fourier Transform}
\newacronym{sde}{SDE}{Stochastic Differential Equation}
\newacronym{nlp}{NLP}{Natural Language Processing}
\newacronym{si}{SI}{Stochastic Interpolant}
\newacronym{rmse}{RMSE}{Root Mean Squared Error}
\newacronym{vs}{VS}{Variogram Score}
\newacronym{lvs}{LVS}{Local Variogram Score}
\newacronym{fm}{FM}{Flow Matching}
\newacronym{fmt}{FMT}{Flow Matching Transformer}
\newacronym{nwp}{NWP}{Numerical Weather Prediction}
\newacronym{ifs}{IFS}{Integrated Forecasting System}
\newacronym{mlp}{MLP}{Multi-Layer Perceptron}
\newacronym{mbm}{MBM}{Member-by-member}
\newacronym{qrn}{QRN}{Quantile Regression Network}
\newacronym{est}{EST}{Energy Score Transformer}
\newacronym{ser}{SER}{Spread-Error Ratio}
\newacronym{scs}{ScS}{Schaake Shuffle}
\newacronym{silu}{SiLU}{Sigmoid Linear Unit}

\AtEveryBibitem{
	\ifentrytype{online}{
		\clearfield{doi}
	}{}
}

\title{Generating ensembles of spatially-coherent in-situ forecasts using flow matching}

\author{%
	David Landry\textsuperscript{1}\thanks{
        Corresponding author.}, Claire Monteleoni\textsuperscript{1,2} and Anastase Charantonis\textsuperscript{1}\\
}

\date{\footnotesize\textsuperscript{\textbf{1}}Inria Paris\\ \textsuperscript{\textbf{2}}University of Colorado Boulder \\
\{david.landry,claire.monteleoni,anastase.charantonis\}@inria.fr}

\begin{document}

\maketitle 


\noindent
\textbf{
\textit{Abstract} --- We propose a machine-learning-based methodology for in-situ weather forecast postprocessing that is both spatially coherent and multivariate.
Compared to previous work, our Flow MAtching Postprocessing (FMAP) better represents the correlation structures of the observations distribution, while also improving marginal performance at the stations.
FMAP generates forecasts that are not bound to what is already modeled by the underlying gridded prediction and can infer new correlation structures from data.
The resulting model can generate an arbitrary number of forecasts from a limited number of numerical simulations, allowing for low-cost forecasting systems.
A single training is sufficient to perform postprocessing at multiple lead times, in contrast with other methods which use multiple trained networks at generation time.
This work details our methodology, including a spatial attention transformer backbone trained within a flow matching generative modeling framework.
FMAP shows promising performance in experiments on the EUPPBench dataset, forecasting surface temperature and wind gust values at station locations in western Europe up to five-day lead times.
}

\section{Introduction}
\label{sec:intro}

Numerical and data-driven gridded weather forecasts suffer from systematic biases when compared against surface observations.
This is mainly attributed to their finite resolution: sub-grid-scale phenomena are not well-represented and prevent a good statistical fit between forecasts and observations.
Consequently, postprocessing is often required before in-situ predictions can be integrated in subsequent forecasting products.

A long-standing challenge for such weather forecast postprocessing models is the preservation of internal correlation structures, including spatial and multivariate coherence.
While correcting forecasts for one given location at a time is well studied~\citep{VannitsemStatisticalPostprocessing2021}, sampling the joint state for many spatial locations requires specialized techniques, especially as the problem dimensionality grows.
This research is critical since multiple applications benefit from increased spatial consistency, such as renewable energy production, energy consumption and hydrological forecasting.

Several methods are available to approach this issue.
Copula-based methods such as \gls{ecc}~\citep{SchefzikUncertaintyQuantification2013} and Schaake Shuffle~\citep{ClarkSchaakeShuffle2004} first perform marginal postprocessing, then reintroduce correlation structures using a dependency template~\citep{LakatosComparisonMultivariate2023}.
\gls{mbm} postprocessing~\citep{SchaeybroeckEnsemblePostprocessing2015} is a marginal postprocessing method, that applies bias and spread corrections separately at each location.
It naturally preserves rank correlation structures among the ensemble members by limiting itself to displacement and rescaling of the gridded forecast.
\gls{ecc} and \gls{mbm} share a common limitation in that they cannot introduce new correlation structures in the prediction: they merely restore or preserve correlations that were already present in the ensemble forecast~\citep{WesterhuisIdentifyingKey2020}.
This does not allow the correction of systematic modeling errors caused by the limited resolution of the underlying prediction.

Another approach to consider is the multivariate extensions of quantile mapping methods~\citep{WhanNovelMultivariate2021}.
\citet{CannonMultivariateQuantile2018} use this strategy by iteratively correcting biases along random rotations of the dataset.
The convergence rate of the algorithm is affected by the dimensionality of the problem, which makes them computationally expensive for larger problems.
Optimal transport quantile mapping methods~\citep{RobinMultivariateStochastic2019} also have a high computational cost that limit the resolution with which we can model high-dimensional distributions. 
Consequently, both of these methods have seen use for problems of small dimensionality (<20 variables).

We contrast this with generative deep neural networks.
Since their introduction for image synthesis applications, they routinely sample very large dimensional distributions~\citep{RombachHighResolutionImage2022}.
Early results were obtained with \glspl{gan}~\citep{GoodfellowGenerativeAdversarial2014}, though their training tends to be a delicate exercise.
This was subsequently addressed by Denoising Diffusion models~\citep{HoDenoisingDiffusion2020} and the closely related \gls{fm}~\citep{LipmanFlowMatching2023}.
They function by approximating a vector field that transports a well-known distribution to a target distribution for which we only have samples.
This provides more stable training and better sample quality than \glspl{gan}, although inference costs are increased because the distribution transport must be solved numerically.

These successes were reflected in weather forecasting applications.
\glspl{gan} were used in weather forecast postprocessing for cloud cover~\citep{DaiSpatiallyCoherent2021}.    
Full generative weather forecasting has been achieved using diffusion models~\citep{PriceProbabilisticWeather2025,CouaironArchesWeatherArchesWeatherGen2024}.

Another weather related example is proposed by~\citet{ChenGenerativeMachine2024}, who also perform spatially-coherent multivariate postprocessing to station locations.
This model, which we refer to as the \gls{esgm}, exploits the cross-correlation sensitive \gls{es} as a training loss.
Random draws from a normal distribution are concatenated to the input feature vector.
The model learns to incorporate this random noise to increase forecast spread in a way that optimizes the \gls{es}.
\gls{esgm} requires the training of multiple models with different training seeds to fully model the distribution of observations from the same gridded forecast.

Following this, we propose \gls{fmap}, a weather forecast postprocessing methodology based on the \gls{fm} generative modeling framework.
It jointly models surface temperature and wind gust values for several spatial locations, making it both spatially coherent and multivariate.
\gls{fmap} has several advantages over existing solutions.
The generated samples model the cross-correlations of the observation distribution more closely, while also improving the marginal forecasts at stations.
Because it does not use a correlation template, it is free to learn new dependency structures from the training data.
A single instance of \gls{fmap} is sufficient to generate high-quality postprocessed forecasts of arbitrary size, despite performing postprocessing for multiple lead times.
The soundness of our approach is demonstrated by training it on the EUPPBench dataset~\citep{DemaeyerEUPPBenchPostprocessing2023} to forecast surface temperature and wind gust at 122 locations in western Europe.

The rest of this paper is organized as follows.
First, section~\ref{sec:problem-statement} states the weather forecast postprocessing problem and introduces notation.
Section~\ref{sec:fm} describes \gls{fmap}, from the \gls{fm} generative modeling framework to the spatial attention transformer backbone.
Section~\ref{sec:benchmark-methods} describes the set of baseline methods we will compare against.
This is followed with a description of our experimental benchmark, including dataset and evaluation metrics, in Section \ref{sec:experiments}.
The results are described in Section~\ref{sec:results} and discussed in Section~\ref{sec:discussion}, along with our concluding remarks.

\section{Problem statement}
\label{sec:problem-statement}

We wish to generate an ensemble of multivariate forecasts $\bm{x}_{t}^i \in \mathbb{R}^D$, where $1 
\leq i \leq M$ is the member index, $t$ is a multi-index designating an \textit{initialization-lead-time} pair, and $D$ is the number of forecast dimensions. 
The forecasts are multivariate in the sense of spatial locations and predicted variables, so that $D = K 
\times V$ is the product of the number of spatial locations $K$ and the number of predicted variables $V$.
The generation is conditioned by an ensemble of gridded weather forecasts. 
These gridded forecasts could be the result of a \gls{nwp} or an AI-based weather forecasting model.
They provide conditioning features $\bm{C}_{t} \in \mathbb{R}^{K \times F}$, the most important of which are the raw forecast for our variable of interest $\bm{w}_t^i \in \mathbb{R}^D$.

We aim to generate samples that are coherent in a spatial and multivariate sense.
We simply define this as being a faithful draw from the distribution of observations $\bm{y_t} \sim q(\bm{x}| \bm{C}_t)$, as opposed to being a statistical construct like a conditional expectation or a marginally-calibrated value.
These samples are of course different to what is obtained by a postprocessing with marginal methods.
Furthermore, by better modeling internal correlation structures, coherent forecasts facilitate their exploitation by downstream forecasting tasks, such as hydrological forecasting, power consumption forecasting, etc.

\section{Method}
\label{sec:fm}

\begin{figure*}
    \includegraphics[width=\textwidth]{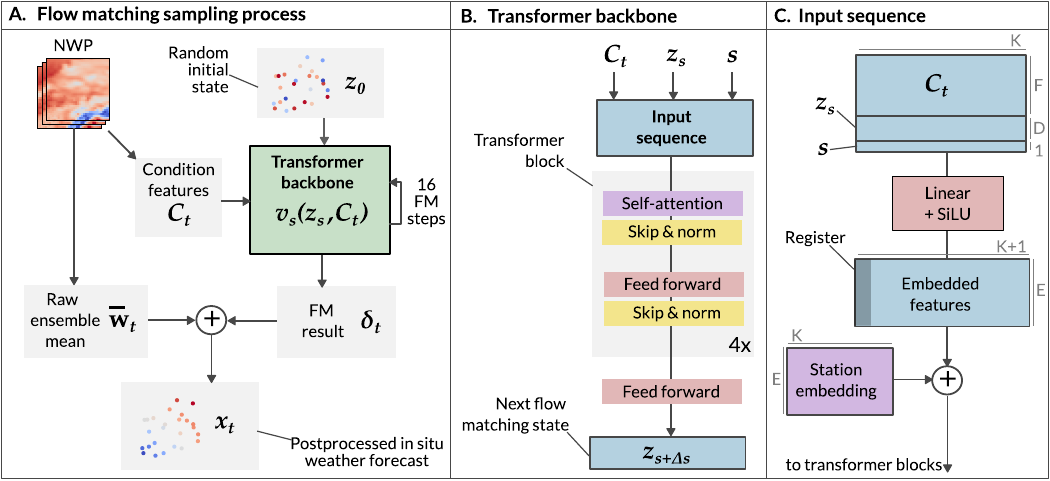}
    \caption{\textbf{A)} The flow matching generation process uses a transformer backbone to iteratively turn an easily-sampled random state into postprocessed in situ forecast.
    Rather than predicting the desired state directly, the generative process predicts the residual from the raw ensemble mean.
    \textbf{B)} Transformer architecture producing the next flow matching state.
    The predictions are made using conditioning features from the underlying forecast $\bm{C}_t$, the previous flow matching state $\bm{z}_s$, and the flow matching time $s$.
     \textbf{C)} Input sequence construction.
     The input values are concatenated together.
     The result is further processed with a linear mapping and a station embedding before being dispatched to the transformer blocks.
     The grayed-out symbol describe the size of the data dimensions.
    }
    \label{fig:big-schema}
\end{figure*}

This section describes our proposed approach, \gls{fmap}, in three steps.
We first introduce the flow matching generative modeling framework.
Then, we state how we use it for weather forecast postprocessing.
We conclude with a presentation of the spatial attention transformer backbone.
The \gls{fmap} implementation used in our experiments is summarized in Figure~\ref{fig:big-schema}.

\subsection{Generation via flow matching}
\label{ssec:flow_matching}

Flow matching~\citep{LipmanFlowMatching2023,LipmanFlowMatching2024} is a generative modeling framework where a model learns how to push a well-known distribution $p(\bm{z})$ towards a target distribution of observations $q(\bm{z})$.
A standard normal distribution is a natural choice for $p$.

\begin{figure}
    \centering
    \includegraphics[width=3.2in]{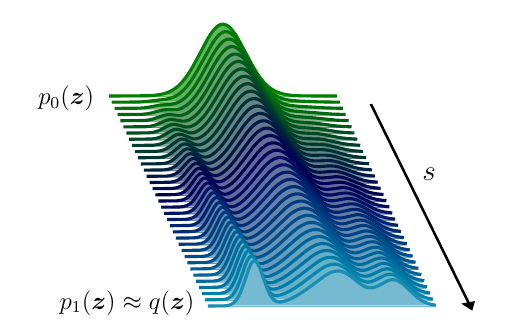}
    \caption{Flow matching starts from a known distribution $p_0(\bm{z})$ to build an approximation $p_1(\bm{z})$ of target distribution $q(\bm{z})$.
    The process takes place during flow matching time $s$.
    }
    \label{fig:distribution-transport}
\end{figure}

The push is done by a flow $\bm{\psi}_s(\bm{z})$ that defines a random variable $\bm{Z}_s$ for any flow matching time $s \in [0,1]$ such that
\begin{align}
    \bm{Z}_s = \bm{\psi}_s(\bm{Z}_0) \sim p_s(\bm{z})
\end{align}
with boundary conditions
\begin{align}
    p_0(\bm{z}) &= p(\bm{z}) \\
    p_1(\bm{z}) &= q(\bm{z}).
\end{align}
This process is illustrated in Figure~\ref{fig:distribution-transport}.

Training a model to predict the flow directly would require full simulations during training, which is impractical.
Fortunately, it is possible to learn a vector field $\bm{v}_s(\bm{z}; \bm{\theta})$ with trainable parameters $\bm{\theta}$ that \textit{defines} the flow, giving us
\begin{align}
    \dfrac{d\bm{\psi}}{ds} &= \bm{v}_s(\bm{z}; \bm{\theta}) \\
    \bm{\psi}_0(\bm{z}) &= \bm{z}.
\end{align}
We generate a flow that respects our constraints by optimizing the vector field using loss
\begin{align}
    \mathcal{L}(\bm{ \theta } ) = \mathbb{E}_{s,p\bm{(z}_0),q(\bm{z}_1)} \| &\bm{v}_s(\bm{\psi}_s(\bm{z}_0|\bm{z}_1); \bm{\theta}) -  (\bm{z}_1 - \bm{z}_0)  \|^2
\end{align}
with
\begin{align}
    \bm{\psi}_s(\bm{z}|\bm{z}_1) = (1 - s)\bm{z} + s\bm{z}_1.
\end{align}
Notice that $\bm{\psi}_s(\bm{z}|\bm{z}_1)$ is the flow conditioned by a target sample $\bm{z}_1$.
Optimizing for it is equivalent to optimizing for the full flow $\bm{\psi}_s(\bm{z})$, but allows us to train the model sample by sample.




Of course our problem is heavily conditioned by the underlying gridded forecast.
Consequently, we train $\bm{v}_s$ to make its predictions given $\bm{C}_t$.

Training a flow matching model involves the following procedure.
To build a training example, we sample a random $\bm{x}_0$ (from a standard normal distribution), a $\bm{C}_t, \bm{x}_t $ couple (from the dataset), and a value of $s$ (different distributions are appropriate, see below).
Secondly, we perform a forward pass to compute loss $\mathcal{L}$, then backpropagate.
Finally, after training, we begin from standard normal samples, then integrate $\bm{v}_s(\bm{z},\bm{C}_t; \bm{\theta})$ over $s$ using a numerical solver.


Flow matching resembles the popular family of diffusion approaches~\citep{SongMaximumLikelihood2021}.
The formalisms used to derive the methods differ, but there exists strong theoretical relationships between the two.
For a more complete introduction to these relationships, and flow matching in general, we refer the reader to~\citet{LipmanFlowMatching2024}.

\subsubsection{Flow matching time sampling during training}
\label{ssec:time-sampling}

To sample $s$ during training, a uniform distribution over $[0, 1]$ is a natural option.
However, one can modify how $s$ is sampled to effectively weight the training loss towards certain regions of the flow matching process.
We use 
\begin{align}
    s = \frac{1}{1+e^{-z}}
\end{align}
with $z \sim \mathcal{N}(0,1)$ for that purpose.
Such a reweighting was empirically shown to improve flow matching results~\citep{EsserScalingRectified2024}, suggesting that properly modeling vector field $\bm{v}_s(\bm{x})$ for central values of $s$ is critical for successful generation.

\subsection{Flow matching for weather forecast postprocessing}
\label{ssec:problem-statement}

The generation procedure for weather forecast postprocessing is depicted Figure~\ref{fig:big-schema}a.
We obtain an in-situ forecast member with sum
\begin{align}
    \bm{x}_t^i = \bar{\bm{w}}_t + \bm{\delta}_{t}^i
\end{align}
where $\bar{\bm{w}}_t$ is the raw forecast ensemble mean.
Forecast residual $\bm{\delta}_t^i$ is the result of the vector field numerical integration
\begin{align}
    \bm{\delta}_t^i = \bm{z}_0^i + \int_0^1 \bm{v}_s(\bm{z}_s^i, \bm{C}_t)ds
\end{align}
with $\bm{z}_s^i$ the flow matching trajectory of the $i$th postprocessed member.
Since the $\bm{z}_0^{i=1..M}$ are all distinct standard normal samples, we obtained spread-out values of $\bm{\delta}_t$.
Starting from the ensemble mean makes intuitive sense, since we expect forecasts $\bm{x}_t^i$ to be closer to $\bar{\bm{w}}_t$ than $\bm{0}$.
This is intended to simplify the distribution transport problem and reduce the number of numerical integration steps at sampling time.

We use a single backbone to postprocess all lead times, since previous results suggested this increases overall performance for neural network models by increasing the amount of training data~\citep{LandryLeveragingDeterministic2024}.
This implies that the \gls{fm} model will have to operate at multiple scales of uncertainties, i.e. the amplitude of a typical $\bm{\delta}_t$ grows with lead time.
To preserve scale invariance in the neural network, we rescale the \gls{fm} output according to the scale of typical model errors.
Our forecast then becomes
\begin{align}
    \bm{x}_t^i = \bar{\bm{w}}_t +\bm{\lambda}_t \odot \bm{\delta}_{t}^i 
\end{align}
where $\bm{\lambda}_t$ is a scaling factor for the lead time and $\odot$ the element-wise product.
The values of $\bm{\lambda}_t$ are chosen via linear regression.
For each variable, the linear model approximates how the raw model error standard deviation grows with lead time.
The linear regression weights are shared across stations.

\subsection{Spatial attention transformer backbone}
\label{ssec:transformer}

Our flow matching backbone, used to predict $\bm{v}_s$, is based on a transformer architecture.
Transformers were initially introduced to address the text translation problem in Natural Language Processing~\citep{VaswaniAttentionAll2017}.
They subsequently proved an appropriate architecture for computer vision tasks~\citep{DosovitskiyImageWorth2021} and full weather forecasting~\citep{BiAccurateMediumrange2023, PriceProbabilisticWeather2025}.
We call our implementation a spatial attention transformer because its self-attention layers are made to operate across spatial locations.
We propose that this is an appropriate representation for this problem: by letting the model transmit information from station to station, the attention layers allow better enforcement of spatial consistency.

\subsubsection{Transformer architecture}
\label{ssec:transformer_architecture}

Our transformer implementation is illustrated in Figure \ref{fig:big-schema}b.
At the top, an input sequence of $K$ tokens is built from the conditioning features $\bm{C}_t$, the flow matching state $\bm{\delta}_s$ and the flow matching time $s$.
Each token in the sequence represents a station individually.

This is followed by a series of transformer blocks, characterized by their self-attention layers.
After this processing is completed the tokens are turned into the next flow matching state using a feed-forward network (containing a sequence of a linear layer, a SiLU activation, and the final linear layer).
We refer the reader to \citet{VaswaniAttentionAll2017} for a more detailed description of the architecture.

\subsubsection{Building the input sequences} 
\label{ssec:sequences}

Our implementation of the architecture being relatively standard, most of our effort is spent designing the input sequences to be processed by the transformer. 
Figure \ref{fig:big-schema}c depicts this process.

We create conditioning features matrix $\bm{C}_t$ by performing nearest-neighbor interpolation between the gridded forecast and the station locations, giving a $K$-wide matrix.
Furthermore, we do not pass all raw ensemble members as conditioning features, but summarize them with their mean and standard deviation across members.
To this we add other metadata features such as the lead time and geographical location.
The combination of all these components gives us an $F$-long feature vector per spatial location.

We concatenate conditioning features $\bm{C}_t$, flow matching state $\bm{\delta}_s$ and flow matching time $s$ (repeated) to form one input features vector for each station.
This is dispatched through a linear layer and an activation layer to form the station tokens.
To this, we append a register token.
Finally, a station embedding is added to the tokens before the whole sequence is sent to the transformer proper.

Our transformer has a dense output, in the sense that we are interested in every output token.
In the absence of special tokens such as ViT's \texttt{[CLS]}, transformers sometimes repurpose spatially meaningful tokens to encode global information~\citep{DosovitskiyImageWorth2021,DarcetVisionTransformers2024}.
To allow aggregated representations inside the transformer, we add a register token that has no spatial meaning to the sequence.
The content of that token is discarded at the output of the transformer.

The transformer is unaware of the tokens spatial relationships.
Consequently it is common practice to inject spatial information in the input sequence~\citep{VaswaniAttentionAll2017}.
To do so we add an embedding $\bm{E} \in \mathbb{R}^{K \times L}$ to the tokens~\citep{DosovitskiyImageWorth2021} immediately after the input dimensionality is expanded to the embedding size $L$.
These embeddings are initialized randomly before training.
Matrix $\bm{E}$ is effectively a station embedding, where the network encodes station characteristics that are relevant for postprocessing.

\section{Baseline methods}
\label{sec:benchmark-methods}

We consider a varied ensemble of baseline methods to compare the proposed models performance against.
We include marginal postprocessing methods to emphasize the effect of modeling internal correlation structures on the generated forecasts.
We include existing generative postprocessing methods to illustrate the improvements brought by \gls{fmap}.

\subsection{Debiased IFS}
\label{ssec:debiased-ifs}

A natural baseline for any weather forecast postprocessing methodology is the uncorrected underlying \gls{nwp} forecast.
Comparing against the raw input gives an approximation of the lift in accuracy brought by postprocessing.
We elect to use the raw \gls{ifs} predictions as our first baseline, with one modification.
Since we produce surface temperature outputs, systematic biases can be caused by differences between station elevation and model elevation at the nearest gridpoint.
These differences are fairly consistent and can be removed through a lapse rate correction.
Rather than performing this correction manually, we have a slightly more flexible approach where we determine prediction biases from data using the climatological periods defined in Section \ref{ssec:data-prep}.
Our debiased \gls{ifs} baseline consists in raw \gls{ifs} forecasts with these biases removed.

\subsection{Distribution Regression Network}
\label{ssec:drn}

The \gls{drn} model is a \gls{mlp} that predicts the parameters of a normal distribution representing the target observations~\citep{RaspNeuralNetworks2018}.
Since its introduction, the \gls{drn} has shown robust results for a variety of weather forecast postprocessing tasks.

Given the conditioning features  $\bm{c}_{t,k}$ related to forecast dimension $1 \leq k \leq D$, the \gls{mlp} is tasked with predicting four parameters $a,b,c,d$ per output dimension.
These are used to construct a normal distribution such that 
\begin{align}
    x^i_{t,k} \sim \mathcal{N}(a + b \bar{w}_{t,k}, e^{c + d \log \sigma_{t,k}}) 
    \label{eqn:emos}
\end{align}
where $\bar{w}_{t,k}$ and $\sigma_{t,k}$ are respectively the mean and standard deviation of the $w^{i=1..M}_{t,k}$.
We apply an exponent on the standard deviation term to preserve positivity during training.

The model is optimized using the \gls{crps}.
The conditioning features are summarized by computing their mean and standard deviation across members before passing them to the \gls{mlp}.

The station embedding is a notable characteristic of the \gls{drn}.
Implementations vary~\citep{RaspNeuralNetworks2018,LandryLeveragingDeterministic2024}, but the general strategy is to reserve a set of trainable weights to represent station identity.
This lets the network register station-specific notes on how to perform postprocessing.
In our case, the station embedding has the same size as the \gls{mlp}s hidden layers.
We add the embedding to the latent features immediately after the first linear layer.

We train one \gls{drn} that makes predictions for all lead time, by providing the lead time as an input feature to the network.

\subsection{Quantile Regression Network}
\label{ssec:qrn}

The \gls{drn} is a flexible approach in the sense that there is no strong coupling between the neural network and how the distribution is represented at the output~\citep{SchulzMachineLearning2022}.
We use this property to add a \gls{qrn} baseline, which replaces the normal distributions of the \gls{drn} with a set of quantiles.
Instead of predicting normal distribution parameters for every output dimension, the network outputs $M$ values per dimension, representing quantile values of the predicted distribution for the observation $y_{t,k}$.
The \gls{qrn} is trained using the \gls{crps} loss, which is equivalent to training it for the quantile loss~\citep{BrockerEvaluatingRaw2012}.
It has a station embedding similarly to the \gls{drn}.
Similarly to its distributional counterpart, we train one \gls{qrn} to predict all lead times.

\subsection{Ensemble Copula Coupling}

A popular way to model spatial correlations is to perform a two-step process where we 1) calibrate the marginal distributions on every dimension 2) recreate rank orderings using a correlation template.
\gls{ecc} and Schaake Shuffle do this by tapping into the underlying gridded forecast and climatology, respectively.
As a representative of these methods we introduce \gls{ecc} in its quantile variant (ECC-Q).
\citet{LakatosComparisonMultivariate2023} present several variants of the method, but state that the widely-used ECC-Q constitutes a powerful benchmark.

We apply \gls{ecc} on the \gls{drn} and the \gls{qrn}.
For the \gls{drn}, we obtain calibrated samples $\tilde{x}_{t,k}^i$ by sampling uniformly spaced quantiles. 
Given the quantile function $F^{-1}_{t,k}(\tau)$ suggested by the predicted normal distribution, we have 
\begin{align}
    \tilde{x}^i_{t,k} = F_{t,k}^{-1}\left(\frac{i}{M+1}\right)
\end{align}
 with $1 \leq i \leq M$.
For the \gls{qrn}, since its output directly models the quantile function of the marginal distributions, the $\tilde{x}_{t,k}^i$ are the network output used as is.

In both cases, we can now generate \gls{ecc} member $x^i_{t,k}$ using the calibrated and ordered forecast members $\tilde{x}_{t,k}^i$ such that
\begin{align}
    x_{t,k}^{i} = \tilde{x}_{t,k}^{\pi(i)}.
\end{align}
Permutation $\pi(i)$ is the rank of $w^i_{t,k}$ across raw ensemble members.

\subsection{Member-by-member neural network}

\gls{mbm} postprocessing~\citep{SchaeybroeckEnsemblePostprocessing2015} is closely related to other spatially-coherent methods.
Because it limits itself to only position and scale adjustments, it preserves the rank orderings in the underlying forecast, while the same rank orderings are restored a posteriori by \gls{ecc}.

A \gls{mbm} model predicts a trio $\alpha_{t,k}$, $\beta_{t,k}$ and $\gamma_{t,k}$ such that
\begin{align}
    x^i_{t,k} = \alpha_{t,k} + \beta_{t,k} \bar{w}_{t,k} +  \gamma_{t,k} (w^i_{t,k} - \bar{w}_{t,k}).
\end{align}
As suggested by \citet{LerchEnhancingMemberbymember2024}, we train a \gls{mlp} to predict these parameters parameters given raw forecast at corresponding location $\bm{c}_{t,k}$.
One such \gls{mbm} model is trained to cover all lead times.

Despite \gls{mbm} predictions being independent for each spatial location, we optimize for the \gls{es}.
This is achieved by making a prediction for each station before backpropagation.
This should improve spatial coherence because the metric is sensitive to the quality of the correlation structures.

\subsection{Energy Score Generative Model}

\citet{ChenGenerativeMachine2024} propose the \acrfull{esgm}, a generative in situ weather forecast postprocessing model that creates varied samples by optimizing the \gls{es}.
The principle of operation is to concatenate a standard normal sample $\bm{z}_{t,k}$ the conditioning features $\bm{c}_{t,k}$.
Since the network is optimized for the \gls{es} (which is sensitive to dispersion and correlation structures), the model learns not to ignore the noise input, and instead uses it to apply dispersion.

The architecture has three networks, respectively used to process the output variables ensemble mean, the output variables ensemble standard deviation, and the conditioning data. 
The first model is linear, while the latter two are \glspl{mlp}.
The network trio is duplicated for each output variable.
It is called iteratively for all spatial locations, output ensemble members and output variable in order to generate a full multivariate realization.
We refer the reader to the original publication for a complete description of the architecture and sampling process.

\citet{ChenGenerativeMachine2024} report that the \gls{esgm} accuracy is improved by training an ensemble of models with different random initializations, and splitting the task of generating an ensemble among them. 
The size of the model ensemble becomes a compromise between forecast accuracy and computational costs.




\section{Experiments}
\label{sec:experiments}

This section describes the experimental benchmark we use to demonstrate the efficacy of \gls{fmap}.
We first describe the dataset, the features we use as input to the different models, and how that data is prepared for input into the neural networks.
We then describe the training and evaluation procedures, including evaluation metrics.

\subsection{Dataset}

We perform our experiments using the EUPPBench dataset~\citep{DemaeyerEUPPBenchPostprocessing2023}.
It consists in paired 0.25\degree gridded forecasts and surface observations from 122 stations in western Europe.
The gridded data are cropped tightly around the station locations, resulting in a $33 \times 32$ grid.

The gridded data contains 11-member bi-weekly reforecasts spanning years 1977-2017, as well as  51-member daily forecasts for years 2017 and 2018. 
They amount to 4180 reforecasts and 730 forecasts.
In both cases the lead times reach up to 5 days, in 6 hour steps for a total of 20 lead times.

EUPPBench provides numerous variables at each grid point which stem from \gls{nwp} model output.
Instantaneous variables include fields surch as surface temperature, while processed variables make 6-hour aggregations (10m wind gust, total precipitation).
Single-level fields are provided, as well as fields for 850, 700 and 500 hPa pressure levels.
We used all data provided by EUPPBench, excluding the Extreme Forecast Indices.
This results in 30 input fields per grid point, including the two fields we are interested in postprocessing (surface temperature and wind gust).
Table~S1 contains the exhaustive list of features used.

The dataset has missing observations over its 20 years span, notably for the wind gust field.
This is typical of in situ observational datasets.
Removing all examples with at least one missing observation from training would have discarded too many examples for our application.
To address this, we remove forecasts with missing observations from the prediction vector $\bm{x}^i_t$ during evaluation.
During training, we rather set the loss related to these predictions to zero, in order to preserve the output shape of the trained models.

We split EUPPBench into a training, validation and test set.
The reforecasts are used for training, except those initialized on years 2003, 2010 and 2016 which are retained for validation.
The 51-member forecasts are used as a test set.
The first three months are removed from the test set and were used for calibration of early generative models.

The original publication for EUPPBench contained results for numerous stations in Switzerland.
Unfortunately these observations could not be distributed freely with the rest of the data and were excluded from the present study.

\subsection{Data preparation and rescaling}
\label{ssec:data-prep}

To preserve positivity of the wind gust field and bring it closer to a standard normal variable, we train the network to predict $\log (1 + x)$ rather than its immediate value (both on the input and output side).

To smoothen the effect of seasonality and the diurnal cycle on our model, we train it to predict anomalies rather than values in natural units.
This treatment also scales the predicted variables around their typical variability, which we posit is beneficial during training.

Given an initialization-lead-time pair $t_\text{ref}$, we define a climatological period $\mathcal{P}_{t_\text{ref}}$ with length $R$ over the training set. 
It contains all forecasts $\bm{w}^i_t$ that 1) have the same initialization hour as $t_\text{ref}$; 2) have the same lead time; and 3) are initialized within 10 days before or after $t_\text{ref}$.
That rolling window size was deemed a good balance between representing the seasonal cycle accurately and smoothing statistical noise in the dataset.
Given $\mathcal{P}_{t_\text{ref}}$ we can compute
\begin{align}
    \mu_{t_\text{ref},k} &= \frac{1}{R}\sum_{t=1}^{R} y_{t,k}  \\
    \sigma_{t_\text{ref},k}^2 &= \frac{1}{R - 1} \sum_{t=1}^{R} (y_{r,k} - \mu_{{t_\text{ref}},k})^2
\end{align}
which we use to rescale model postprocessing model output $\tilde{x}_{t,k}^i$
\begin{align}
    x_{t,k}^i = \sigma_{t,k} \tilde{x}_{t,k}^i + \mu_{t,k}.
\end{align}
We perform an analogous conversion for the input using model climatology instead of observation climatology.

The conditioning features $\bm{C}_t$ warrant some preprocessing as well, but typically do not require a procedure quite as involved.
Instead, they are scaled by their mean and standard deviation over the training set so that they are roughly standard normal.
As mentioned in Section \ref{ssec:sequences}, these features are summarized by computing their first and second moment across members.
Consequently the number of input features is doubled.

This data preparation procedure was applied systematically to all benchmark models as well as \gls{fmap}.

\subsection{Model implementations}

The \gls{drn}, \gls{qrn} and \gls{mbm} models are configured with four hidden layers.
The embedding size is set to 256 and SiLU activation functions are used.
The \gls{qrn} outputs 51 quantile values to match the ensemble size of the test set.
These values are largely inspired from preceding studies~\citep{LandryLeveragingDeterministic2024}.

We reimplemented the \gls{esgm} for this work. 
To better align the \gls{esgm} to other baselines, we applied some modifications to it, all of which improved validation scores on our benchmark.
We modify the architecture to add a station embedding after the first layer of the conditioning data \gls{mlp}.
Every \gls{esgm} instance is trained on all lead times to maximimze dataset size.
Furthermore, we increase the size of the \glspl{mlp} to four hidden layers with an embedding size of 512.
The other hyperparameters (size of random feature vector $\bm{z}_{t,k}$, number of model instances, size of ensembles used to train the model) were kept at their original values (respectively 10, 10 and 50).

For \gls{fmap}, we configure the transformer with four attention blocks, having four attention heads each. 
The embedding size is set to 1024.
We use a dropout rate of 10\%.
In the feed-forward networks at the end of the attention blocks, the internal representation size is kept constant.
At sampling time, we perform numerical integration using Euler's method with uniform step sizes.
The number of steps is set to 16 unless otherwise specified.

\subsection{Training}

All models are trained using the AdamW optimizer and the PyTorch \texttt{OneCycle} learning schedule.
\gls{fmap} is trained with a maximum learning rate of $10^{-4}$ over 80 epochs.
The \gls{mlp}-based models (\gls{drn}, \gls{qrn}, \gls{mbm}, \gls{esgm}) are trained with a maximum rate of $10^{-3}$ over 50 epochs.

\subsection{Evaluation}

This section describes the suite of evaluation metrics used throughout our study, starting with dimension-wise evaluation, before covering multivariate evaluation metrics.

\subsubsection{CRPS}

The \gls{crps}~\citep{GneitingStrictlyProper2007} is a proper scoring rule that is widely used to the evaluation of probabilistic forecasts.
Given single-dimensional ensemble forecasts $X_{t,k} = x_{t,k}^{i=1..M}$ for output dimension $k$, we compute the \gls{crps} against corresponding observation $y_{t,k}$ using its empirical estimator
\begin{align}
    \text{CRPS}(X_{t,k}, y_{t,k}) =\frac{1}{M}&\sum_{i=1}^N | x_{t,k}^i - y_{t,k} | \notag\\
        &- \frac{1}{2M^2} \sum_{i,j=1}^N | x_{t,k}^i - x_{t,k}^j |,
\end{align}
where $| \cdot |$ is the absolute value.

Being univariate, the \gls{crps} does not help us evaluate how model can reconstruct correlation structures (spatially and across variables).
However, it is easily interpretable because it is expressed in the natural units of the forecast.

\subsubsection{Brier Score}

The \gls{bs} is another marginal evaluation tool at our disposal, focused on forecast accuracy for extreme values.
Its exceedance thresholds are computed separately by spatial location.
Threshold $Y^\tau_{t,k}$ is the $\tau$-quantile of the observation dataset, within the climatological period defined in Section~\ref{ssec:data-prep}.
We can then compute the \gls{bs} via
\begin{align}
    \text{BS}_\tau(X_{t,k}, y_{t,k}) = \Big( \mathbf{1}[y_{t,k} > Y^\tau_{t,k}] - \frac{1}{M} \sum_{i=0}^M \mathbf{1} [x^i_{t,k} > Y^\tau_{k}] \Big)^2
\end{align}
where $\mathbf{1}[\cdot]$ is the indicator function.

\subsubsection{Spread-error ratio}

By assuming exchangeability between all ensemble members and the observation, one can derive a relationship between a models ensemble mean \gls{rmse} and typical ensemble spread~\citep{FortinWhyShould2014}. Given 
\begin{align}
    \text{Spread} &= \sqrt{\frac{1}{T}\sum_{t=1}^T \frac{1}{M-1} \sum_{i=0}^M (x^i_{t,k} - \bar{x}_{t,k})^2} \\
    \text{Error} &= \sqrt{\frac{1}{T} \sum_{t=1}^T (\bar{x}_{t,k} - y_{t,k})^2} 
\end{align}
we get spread-error ratio
\begin{align}
\text{SER} = \sqrt\frac{M+1}{M}\frac{\text{Spread}}{\text{Error}}    
\end{align}
which is a widely used metric in forecast verification to assess model dispersivity.
This verification tool does not apply to postprocessing models for which we are not willing to make exchangeability assumptions, like models predicting quantiles.

\subsubsection{Energy Score}
\label{ssec:es}

The \gls{es} is a multi-dimensional extension of the \gls{crps}.
It allows evaluating the spatial and multivariate consistency of the $D$-dimensional forecast, making it especially useful for this study.
Given an ensemble forecast $\bm{X}_t = \bm{x}_t^{i=1..M}$ and the corresponding observation $\bm{y}_t$, we compute the \gls{es} using its empirical formulation
\begin{align}
    \text{ES}(\bm{X}_t, \bm{y}_t) = \frac{1}{M} \sum_{i=1}^M \| \bm{x}^i_t - \bm{y}_t \| - \frac{1}{2M^2} \sum_{i,j=1}^M  \|\bm{x}^i_t - \bm{x}^j_t \|,
    \label{eqn:es}
\end{align}
where $\| \cdot \|$ is the euclidean norm.

The \gls{es} is also a proper scoring rule, though its sensitivity to misrepresentation of internal correlation structures is being discussed~\citep{PinsonDiscriminationAbility2013,ZielMultivariateForecasting2019}.
Nevertheless, it is worthwhile to add other metrics that evaluate the quality of multivariate dependencies.

\subsubsection{Variogram Score}
\label{ssec:variogram-score}

The \gls{vs} measures how well the internal correlation structures of the data are represented~\citep{ScheuererVariogramBasedProper2015}.
It is not sensitive to simple biases, only to the intervariate correlations.
This is both a blessing and a curse. The \gls{vs} cannot be used on its own, because it could miss simple biases,  but it allows us to study cross-correlation errors in isolation~\citep{DaiSpatiallyCoherent2021}.

Given an ensemble forecast $\bm{X}_t = \bm{x}^{i=1..M}_t$ for a $N$-dimensional observation $\bm{y}_t$, the \gls{vs} is computed using
\begin{align}
    \text{VS}(\bm{X}_t, \bm{y}_t) =  \sum_{i,j=1}^N  \left( |y_{t,i} - y_{t,j}|^\rho - \frac{1}{M} \sum_{m=1}^M | x_{t,i}^m - x_{t,j}^m |^\rho \right)^2.
\end{align}
Parameter $\rho$ is set to $\frac{1}{2}$ following \citet{ScheuererVariogramBasedProper2015}.

A limitation of the \gls{vs} is that it loses sensitivity in the presence of strongly uncorrelated variables.
This is noticeable for spatially distant stations where the observations are weakly correlated.
One can mitigate this empirically by weighing the score with the inverse of the station mutual distance~\citep{ScheuererVariogramBasedProper2015}.
Alternatively, we can use a procedure where the \gls{vs} is computed locally around stations, rather than the full collection of spatial locations~\citet{ChenGenerativeMachine2024}.
We choose the latter and define a \gls{lvs}.
It consists in computing, for all stations, the \gls{vs} of the model for the $K$ nearest stations neighborhood, meaning evaluation is $K$-dimensional.
In some cases we also evaluate a multivariate version of this metric, where both surface temperature and wind gust speed are included in the evaluation (yielding a $2K$-dimensional evaluation).
We set $K=5$ throughout this study, a value that allows comparison with previous work~\citep{ChenGenerativeMachine2024}.

\subsubsection{Power spectral density}

Early AI-based weather forecasting models exhibited blurry forecasts because they are trained to predict the conditional expectation of the distribution rather than an actual realization~\citep{BonavitaLimitationsCurrent2024}.
To control for such deficiencies it is common to evaluate the power spectral density of a models predictions.
By showing how different frequencies are represented in the forecasts, these plots allow us to diagnose under-representation of high frequencies.

This type of analysis is typically done on gridded forecasts.
In that spirit we bring our point-wise forecasts back on a 0.1\degree grid for this evaluation.
The 0.1\degree  resolution is convenient because each station uniquely maps to its nearest gridpoint.
To construct the grid, we start from an all-zero field, then place the predicted anomaly values for each station on their corresponding gridpoint.
To avoid high-frequency artifacts stemming from the construction of this grid (as we transition from the background to gridpoints where stations are present), we apply a gaussian convolution on the grid before computing the power spectrum densities.
An example grid construction can be viewed in Figure~S1.


\subsubsection{Skill Scores}

To facilitate the interpretation of some figures, we compute metrics in terms of their corresponding skill scores.
Given a metric $\bar{S}$ aggregated for a model over the test set, its skill score $SS$ is
\begin{align}
    SS = 1 - \frac{\bar{S}}{\bar{S}_{baseline}},
\end{align}
where $\bar{S}_{baseline}$ is the score of an appropriate baseline.
This skill score is interpreted as a percentage improvement/degradation over the baseline.
As a baseline we typically use the \gls{drn} model from Section \ref{ssec:drn} with \gls{ecc}.

\subsubsection{Statistical significance test}
\label{ssec:bootstrap}

Where we desire estimating the statistical significance of our results, we use the pairwise bootstrap procedure described by \citet{HamillHypothesisTests1999} with 500 bootstraps and 5\% to 95\% confidence intervals.

\section{Results}
\label{sec:results}

We begin this section with an evaluation of the models capability to perform spatially coherent multivariate weather forecast postprocessing.
In a second step, we evaluate their univariate performance.
Then, we plot power spectra for representative models, which is indicative of how close the forecast members are from the distribution of observations.
We conclude this section with a single forecast case study, and and an assessment of how \gls{fmap} behaves under different scaling scenarios.

\subsection{Spatially-coherent weather forecast postprocessing}
\label{ssec:results-postprocessing}

Table \ref{tab:multivariate_metrics} shows multivariate evaluation metrics computed over the test set for our postprocessing models.
First, we note how \gls{fmap} has the best performance for all metric-variable combination.
Secondly, we underline the remarkable results of the \gls{mbm} neural network, showing that relatively simple models can preserve spatial consistency when trained with the \gls{es}.
Finally, we observe some limitations of \gls{ecc} when coupled with the \gls{drn}, which degrades the local variogram score compared to the debiased \gls{ifs} model.
In line with previous work~\citep{LandryLeveragingDeterministic2024}, all models show reasonable skill when training single model instances for postprocessing at multiple lead times.

\begin{table}
    
\caption{Postprocessing model performance for spatially coherent forecasts.
Values are aggregated for all lead times.
Lower is better. 
The best score is \textbf{bold}, the second best is \underline{underlined}.}
\label{tab:multivariate_metrics}
\footnotesize
\centering
\begin{tabular}{lcccccc}
\toprule
 & \multicolumn{3}{c}{Energy Score} & \multicolumn{3}{c}{Local Variogram Score} \\
Variable & Both & Temp. & Wind & Both & Temp. & Wind \\
\midrule
Debiased & 21.60 & 13.04 & 16.75 & 7.72 & 1.15 & 1.90 \\
DRN-ECC & 18.82 & 11.14 & 14.83 & 8.14 & 1.78 & 2.21 \\
QRN-ECC & 18.60 & 10.92 & 14.67 & 6.99 & 1.19 & 1.75 \\
MBM-MLP & 18.39 & \underline{10.80} & 14.49 & \underline{6.61} & \underline{0.93} & \underline{1.65} \\
ESGM & \underline{18.36} & 10.83 & \underline{14.48} & 7.09 & 1.28 & 1.83 \\
FMAP & \textbf{18.13} & \textbf{10.55} & \textbf{14.30} & \textbf{6.17} & \textbf{0.81} & \textbf{1.44} \\
\bottomrule
\end{tabular}

\end{table}

\begin{figure*}
    \centering
    
    \includegraphics[width=\textwidth]{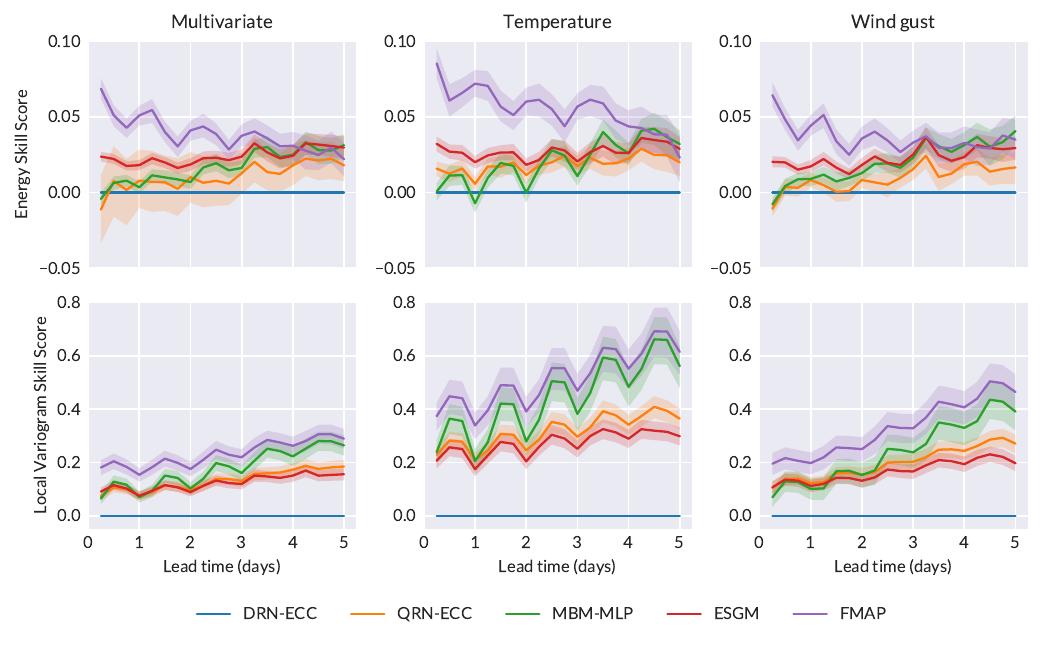}

    \caption{Postprocessing model skill scores for spatially coherent forecasts according to lead time.
    Higher is better.
    The baseline for skill scores is the Distribution Regression Network with Ensemble Copula Coupling (DRN-ECC).
    Shaded areas are the result of a pairwise bootstrap procedure with 5 to 95\% confidence interval.}
    \label{fig:panel_multivariate_metrics}
\end{figure*}

Figure \ref{fig:panel_multivariate_metrics} plots \gls{es} and \gls{lvs} skill for each model.
In both cases the skill score baseline is the \gls{drn} with \gls{ecc}.
The \gls{es} difference between \gls{fmap} and the other models is stronger in early lead times, while the \gls{lvs} shows more consistent improvements.

\subsection{Marginal performance}

Table \ref{tab:metrics_marginal} summarizes station-wise evaluation metrics, aggregated for all lead times.
\gls{fmap} gives best marginal performances, showing its elaborate generative process did not degrade marginal forecasts.
The \gls{qrn} gave good results among our baseline methods.

\begin{table}
\caption{
Weather forecast postprocessing models performance for marginal metrics. 
Values are aggregated for all lead times.
Brier scores are given for the 5th and 95th percentile thresholds.
Lower is better.
The best score is \textbf{bold}, the second best is \underline{underlined}.
}
\label{tab:metrics_marginal}
\centering
\footnotesize

\begin{tabular}{lccccc}
\toprule
& \multicolumn{3}{c}{Surface Temperature} & \multicolumn{2}{c}{Wind Gust} \\
Model & CRPS & \makecell{BS$\times 10^2$ \\ (5th)} & \makecell{BS$\times 10^2$ \\ (95th)} & CRPS & \makecell{BS$\times 10^2$ \\ (95th)} \\
\midrule
Debiased & 0.96 & 1.16 & 3.35 & 1.20 & 3.00 \\
DRN-ECC & 0.80 & 0.96 & 2.72 & \underline{1.04} & 2.68 \\
QRN-ECC & \underline{0.79} & \underline{0.92} & \underline{2.66} & \underline{1.04} & \underline{2.54} \\
MBM-MLP & 0.80 & 0.96 & 2.76 & \underline{1.04} & 2.59 \\
ESGM & \underline{0.79} & 0.93 & 2.68 & \underline{1.04} & 2.58 \\
FMAP & \textbf{0.77} & \textbf{0.90} & \textbf{2.60} & \textbf{1.03} & \textbf{2.50} \\
\bottomrule
\end{tabular}

\end{table}

\begin{figure}
    \centering
    \includegraphics{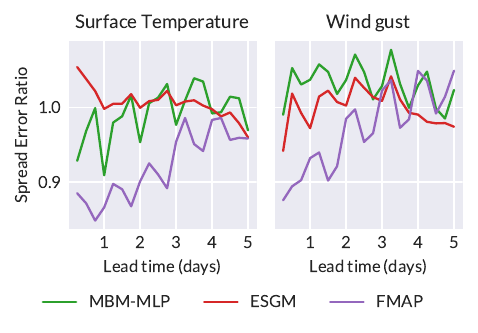}

    \caption{Postprocessing model spread-error ratios.}
    \label{fig:spread_error_per_lead}
\end{figure}

Figure \ref{fig:spread_error_per_lead} shows the spread-error ratio according to lead time. 
The \gls{drn} and \gls{qrn} are excluded since their members are not exchangeable, which is an assumption made when using the spread-error ratio~\citep{FortinWhyShould2014}.
\gls{fmap} is underdispersive at smaller lead times, despite having better marginal metric scores in Table~\ref{tab:metrics_marginal}.
Interestingly, this matches results obtained by some diffusion models for full weather forecasting (though with less intensity).
\citet{CouaironArchesWeatherArchesWeatherGen2024} alleviate this using noise scaling, which consists in increasing the variance of the standard uniform samples used to initiate generation.
We leave such experiments for future work.

\subsection{Spectral properties}

\begin{figure}
    \centering
    \includegraphics{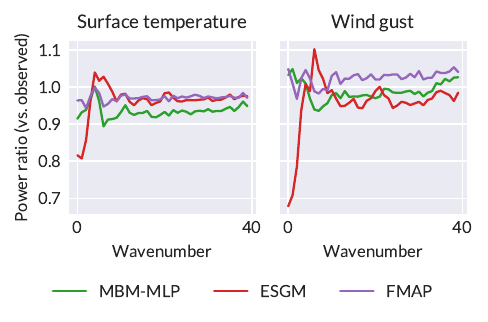}

    \caption{
    Power spectrum density ratio for postprocessing models.
    The ratio is computed against the mean power spectrum of the corresponding observations.
    The densities are averaged on the test set, at 3-days lead time.
    }
    \label{fig:power}
\end{figure}

We plot the power spectrum of different postprocessing models for wind gust fields in Figure \ref{fig:power}.
The plots show power ratio with respect to the spectrum of the observations.
The spectra are computed using anomaly values rather than values in natural units.

\gls{fmap} power signatures match those of the observations well, having a power ratio close to one across the spectrum.
The \gls{esgm} has less energy in the low frequencies, indicating less representation of large scale structures in the maps.

\subsection{Case study}

\begin{figure*}
    \centering
    \includegraphics[width=\textwidth]{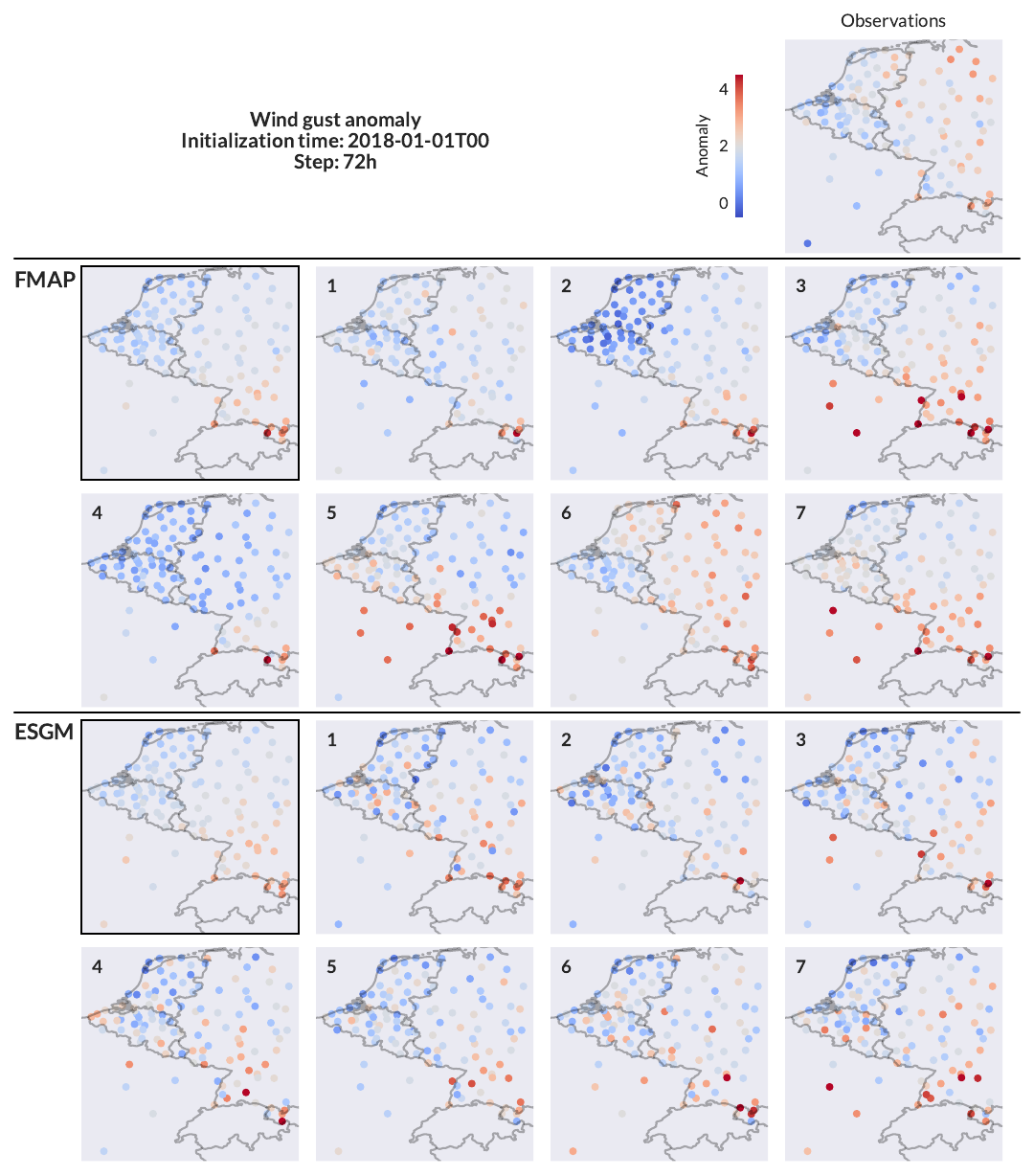}
    \caption{Sample forecasts for our model (FMAP) and the Energy Score Generative Model (ESGM).
    The values are displayed as anomalies according to a rolling window climatology.
    The framed maps represent the mean of the generated ensembles.}
    \label{fig:forecast_fm}
\end{figure*}

We illustrate the benefits of our approach by showcasing \gls{fmap} and \gls{esgm} wind gust forecasts in Figure \ref{fig:forecast_fm}.
The values are displayed as anomalies according to the climatological period defined in Section~\ref{ssec:data-prep}.
\gls{fmap} successfully recreates the peppering of stronger wind gust measurements, similarly to what is visible in the matching observations.
It makes predictions with varied mesoscale configurations (contrast member 2 with member 6, for instance), despite being never conditioned on specific samples from the underlying \gls{nwp} forecast (only ensemble mean and standard deviation).

\subsection{Scaling studies}

Table \ref{tab:scaling} contains results for scaling studies performed on the flow model.
These experiments control for the size of the input forecast, the size of the postprocessing forecast, and the number of steps taken during sampling.

\begin{table}
    \caption{Scaling studies for the proposed methodology. 
    The Energy Skill Scores (ESS) are computed against our standard configuration: 51 input members, 51 output members and 16 sampling steps.
    The scores are aggregated for all lead times.}
    \label{tab:scaling}
    \footnotesize
    \centering    
        \begin{tabular}{lcc}
        \toprule
        \makecell{Parameter} & \makecell{Value}  & ESS \\
        \midrule
        N Members (Input) & 4 & -0.023 \\
        & 8 & -0.008 \\
        & 16 & -0.003 \\
        & 32 & -0.001 \\
        & 51 & 0.000 \\
        \midrule
        N Members (Output) & 2 & -0.895 \\
         & 4 & -0.278 \\
         & 16 & -0.043 \\
         & 32 & -0.012 \\
         & 51 & 0.000 \\
         & 64 & 0.004 \\
         & 128 & 0.012 \\
         & 256 & 0.016 \\
         \midrule
         N Steps & 4 & -0.027 \\
         & 8 & -0.001 \\
         & 16 & 0.000 \\
         & 32 & -0.002 \\
        \bottomrule
    \end{tabular}
\end{table}

\subsubsection{Input members}
Removing members from the underlying gridded forecast moderately reduces performance because the \gls{fm} is conditioned on less accurate estimations of the weather state.

\subsubsection{Output members}

Interpreting performance improvements over varying ensemble sizes requires some care~\citep{LeutbecherEnsembleSize2019}.
To reason about this we use the analogy with the \gls{crps}, which is the univariate version of the \gls{es}.
In the marginal case, when computing the \gls{crps}, a decrease of the error metric is expected when ensemble size increases.
This bias can be compensated using Fair \gls{crps} as proposed by \citet{FerroFairScores2014} when working in one dimension.
To the best of our knowledge,  no multivariate equivalent has been proposed.
Consequently, we rely on the empirical formulation in Equation \ref{eqn:es} for Table~\ref{tab:scaling}, and this has to be kept in mind when assessing the results.

At the very least, the table indicates at least some variability in the generated samples, since a degenerate distribution should be penalized for being overconfident.
The benefits of ``sample resolution'' keep increasing for samples sizes up to 256 in our benchmark.

\subsubsection{Sampling steps}
The proposed flow matching model is computationally more demanding that other methods because it involves multiple neural network calls during inference.
Table~\ref{tab:scaling} shows the effect of reducing the number of sampling steps on the \gls{es}. 
It suggests that less expensive sampling procedures could be considered for the current model.

\section{Discussion and conclusion}
\label{sec:discussion}

In this work we proposed \gls{fmap}, a new weather forecast postprocessing methodology based on a spatial attention transformer and flow matching.
\gls{fmap} achieves state of the art weather forecast postprocessing.
It is both spatially coherent and multivariate: it reflects the cross-correlation structures present in the observations more faithfully than baseline methods.
That is achieved without hindering marginal forecasting performance.
\gls{fmap} is not limited to modeling correlation structures that are present in the underlying forecast --- it can implement new structures inferred from training data.
Our methodology requires training only one model whereas previous work involve training and inferring from multiple random seeds to increase spread. 
Furthermore, it is not limited to the ensemble size of the underlying forecast, and can generate an arbitrary number of samples from one numerical prediction.
Taken together, these properties constitute a step forward in weather forecast postprocessing.

Like any methodology using flow matching or diffusion, our approach suffers from high inference cost.
Sampling one batch of postprocessing forecasts requires several neural network calls.
We argue this is still negligible given the cost of the underlying numerical/AI-based forecast, which we contrast with our models intermediate size (4 attention blocks).
Reducing the number of steps required for sampling flow matching models is an active research area~\citep{LiuFlowStraight2022, EsserScalingRectified2024,SalimansMultistepDistillation2024,YinOnestepDiffusion2024}, suggesting this cost could be further reduced in the future.

The scope of our own study also has limitations, in ways that constitute interesting future work.
We only experimented with fixed step-size Euler solver for sampling.
In other fields, gains were achieved using non-uniform step sizes~\citep{EsserScalingRectified2024} and second degree solvers~\citep{KarrasElucidatingDesign2022}.
We studied surface temperature and wind gust in this work, but precipitation and cloud cover fields are also of high interest for spatially-coherent forecasting.
These fields involve challenging distributions which could require specific adaptations to the framework.
A promising avenue is to adapt flow matching/diffusion framework to better model heavy-tailed distributions~\citep{ShariatianDenoisingLevy2025}.
Encouraging results were recently obtained on weather related applications~\citep{PandeyHeavyTailedDiffusion2024}.
These heavy-tailed distribution could also improve the representation of extreme weather events in general, which is crucial, given our changing climate.
Extreme-oriented studies of the proposed methodology could address the underdispersivity we measure in Figure \ref{fig:spread_error_per_lead}.

An obvious improvement on this work would be to extend the generation to the time axis, to model spatio-temporal correlation structures.
Current diffusion-based weather forecasting neural networks are spatially generative, but autoregressive in the time axis~\citep{PriceProbabilisticWeather2025,CouaironArchesWeatherArchesWeatherGen2024}.
\citet{AndraeContinuousEnsemble2024} propose a to generate temporally-coherent samples by correlating the driving noise across sampling procedures, but show that their method is best used in a hybrid auto-correlated/time-continuous manner.
An underlying issue is the very high dimensionality of gridded trajectories, which is preventing joint spatio-temporal sampling.
This is less of an issue for in situ postprocessing because the output state is an order of magnitude smaller.
Consequently, we could see spatio-temporal generation be developed earlier for postprocessing than for full gridded weather forecasting.

Our case study identified a forecast where modeling topography-rich areas showcased the benefits of our approach. 
We believe a dedicated study over a mountainous area, perhaps improving a higher-resolution \gls{nwp} model, could demonstrate more benefits.

\section*{Acknowledgements}

The authors would like to thank Guillaume Couairon and Emmanuel de Bézenac for their fruitful comments.
This work was supported by a Choose France Chair
in Artificial Intelligence grant from the French government.
It was performed using HPC resources from GENCI-IDRIS (Grant AD011014334).

\section*{Data Availability Statement}

This work was realized using data from the publicly available EUPPBench dataset, available at \url{https://eupp-benchmark.github.io/}.


\printbibliography


\raggedbottom

\clearpage
\onecolumn
\appendix

\end{document}